\documentclass[prb,twocolumn,showpacs]{revtex4}
\begin{document}

\title{Electronic excitations and the tunneling spectra of metallic nanograins}

\author{Gustavo A. Narvaez$\,^{a,b}$ and George Kirczenow$\,^{b}$}

\affiliation{$^{a}\,$Department of Physics, The Ohio State University, 
Columbus, Ohio 43210 \\
%}
%\affiliation
%{
$^{b}\,$Department of Physics, Simon Fraser University, Burnaby,
British Columbia, Canada V5A 1S6}

\date{\today}

%\maketitle

\begin{abstract}

Tunneling-induced electronic excitations in a metallic nanograin 
are classified in terms of {\em generations}: subspaces of 
excitations containing a specific number of electron-hole pairs. 
This yields a hierarchy of populated excited
states of the nanograin that strongly depends on (a)
the available electronic energy levels; and (b)
the ratio between the electronic relaxation rate within
the nano-grain and the bottleneck rate for tunneling
transitions. To study the response of the electronic
energy level structure of the nanograin to the excitations, and its signature
in the tunneling spectrum, we propose a microscopic mean-field theory. 
Two main features emerge when considering an Al nanograin coated with
Al oxide: (i) The electronic
energy response fluctuates strongly in the presence of disorder, from level to level
and excitation to excitation.
Such fluctuations produce a dramatic sample dependence of the tunneling
spectra.  On the other hand, for
excitations that are energetically accessible at low applied
bias voltages, the magnitude of the response, reflected in the renormalization of the single-electron
energy levels, is smaller than the average spacing
between energy levels.  (ii) If the tunneling and electronic
relaxation time scales are such as to admit a significant
non-equilibrium population of the excited nanoparticle
states, it should be possible to realize much higher spectral
densities of resonances than have been observed to date in
such devices. These resonances arise from tunneling into
ground-state and excited electronic energy levels, as well as
from charge fluctuations present  during tunneling.
\end{abstract}

\maketitle
 
\section{Introduction}

In the late '90s it was demonstrated that single quantum level tunneling spectroscopy 
is a powerful tool for studying the physics of simple, noble, and 
magnetic nano-scale metals. Ralph, Black and Tinkham first used this technique to study the
electronic energy level structure of {\em individual} oxide-coated aluminum
nanoparticles\cite{ralph_black_tinkham}, and it was soon applied to Co, Au, Ag, and
Cu nanograins as well\cite{davidovic_PRB_2000,noble,ferromagnetic}.  Even after several
theoretical studies\cite{agam_PRL_1997,theories,narvaez_PRB_2002,narvaez_PRB_2002b,narvaez_PRB_2003}, a 
key aspect of the data (the unexpectedly high density of resonances in the
tunneling spectra of the metal nanograins) remains not fully understood, as was discussed in Ref.
\onlinecite{davidovic_PRB_2000}. More intriguingly, these resonances appear in
clusters. As the nanograins are expected to present strong surface disorder, this bunching of
resonances seemed to  collide with the predictions of random matrix theory for the
single-particle energy levels of  small disordered conductors\cite{brody_RMP_1981}.
In order to explain these experimental features, Agam and co-workers
\cite{agam_PRL_1997} argued that the electron tunneling may occur under conditions far from
equilibrium which would result in the presence of a large number of
resonances and bunching. They offered a phenomenological model able to account for some of the
observed features. More recently, however, Davidovi\'c and Tinkham presented alternate 
scenarios---charge traping, single occupancy of Kramers doublets, and 
non-equilibrium effects---that might be responsible for the bunching of the tunneling 
resonances that they observed in Au nanoparticles\cite{davidovic_PRB_2000}. Furthermore, the
present authors developed a microscopic model  for the tunneling spectroscopy of these
nanoparticles and argued that clustering of resonances should arise naturally in metal
nanograins\cite{narvaez_PRB_2002}, even in the absence of the non-equilibrium transport
effects introduced by Agam {\em et al.}
\cite{agam_PRL_1997} It has also been shown recently that charge-fluctuations
that are present during the electron tunneling should generate {\em additional}
tunneling resonances,\cite{narvaez_PRB_2002b} not considered in previous
theories, and that penetration of environmental electric 
fields into metal nanoparticles can have striking effects on their
tunneling spectra.\cite{narvaez_PRB_2003} 

In this paper, we classify the electronic excitations that may take place during electron tunneling  
within an ultrasmall metallic grain in
terms of {\em generations}: subspaces of excitations with a specific number of electron-hole pairs.
Furthermore, we propose a general microscopic mean-field model to calculate the quasi-particle 
energy levels in the presence of an excitation; such a model has not been
previously presented in the literature, to the best of our knowledge.
The generations form a hierarchy that strongly depends on (i) the number of available
electronic energy levels for tunneling, and (ii) the ratio
between the electronic relaxation rate within the nanograin and the bottleneck
rate for tunneling transitions. The latter is quite sensitive to the thickness of the tunneling
barriers.
The applied bias voltage, and characteristics of the device---grain-lead 
capacitances, charging energy, and typical energy spacing between
electronic energy levels in the nanograin---determine the number of electronic configurations
that are present in a generation.
Finally, we present detailed results of the response of the electronic energy structure to the excitations 
in an ultra-small aluminum grain coated with Al oxide. We find that: (i) The renormalization of
the electronic energy levels  in the ground states fluctuates strongly from level to level, and
excitation to excitation due to the stochasticity of the confined electronic wavefunctions that
is imposed by the disorder present in the grain; these fluctuations depend dramatically on
the specific realization of the disorder. (ii) The average single-electron energy level spacing in the
excited states  remains nearly unchanged when compared with the value obtained in the ground-state;
changes are within a few percent.
 (iii) If the kinetics of the tunneling transitions, and intragrain electronic
relaxation, are such as to admit  a significant population of excited states, then the predicted number
of  tunneling resonances can be {\em much} higher than has been
observed to date in tunneling experiments in non-magnetic nanoparticles. (iv) Disorder is
responsible for a strong sample-dependence of (a) the number of resonances present, as clusters,
in the tunneling spectrum, and (b) the spectral width of such clusters as a function of the applied bias voltage.

%%  mechanism %%
%
\section{Mechanism for electronic excitations and their signature in nonequilibrium tunneling
spectroscopy}
\label{mechanism}

We are interested in the
electron tunneling regime where: (i)  the Coulomb charging energy\cite{CB}
($U$) of the grain that results from adding or removing a valence electron, is greater than
the expected average particle-in-a-box level spacing ($\delta$) around the Fermi energy. 
This is often the experimental situation\cite{ralph_black_tinkham,davidovic_PRB_2000,noble,ferromagnetic}. 
(ii) The electronic relaxation rate ($\Gamma_r$) within the
grain is much {\em smaller} than the bottleneck tunneling rate ($\Gamma$), and (iii)
tunneling leaves the grain with a surplus or deficit of one electron with respect to the $n_0$
valence electrons that make the  grain neutral.
In this regime, after an
electron tunnels in {\em and} out, the grain may be left in an excited
electronic state while still being neutral. Therefore, charge transport
may take place via excited (non-equilibrium) nanoparticle states.
This mechanism for generating electronic excitations, that may be reflected in the energy
spectra of metallic nanograins, was first exploited by Agam, Wingreen, Altshuler,
Ralph and Tinkham\cite{agam_PRL_1997} (AWARTi)
to explain tunneling spectroscopy experiments in Al  nanoparticles. 
However, the original AWARTi model was spinless and, more importantly, the
effects of the electronic excitations on the electronic energy levels were treated only within
the framework of random matrix theory.

Let us now discuss the {\em genesis} of the tunneling-induced electronic excitations.
We begin by considering a two-terminal device consisting of macroscopic source
({\em S}) and drain ({\em D})  electrodes separated by a thin insulating 
layer from a metallic nanoparticle ($d$). Following the
Orthodox theory of Coulomb blockade,\cite{CB} the
electrochemical potential of the source (drain) electrode is set to:
$\mu^{S(D)}(V)=E_F+(-)(C_{D(S)}/C_{\Sigma})eV$; where $V$ is the applied bias
voltage, $e$ the magnitude of the electron charge, and
$C_{\Sigma}=C_{S}+C_{D}$ with $C_{S(D)}$ the capacitance between electrode {\em S}({\em D}) and the 
nanoparticle. 
The quantum nature of the nanoparticle is taken into account
through its discrete electronic structure. For definiteness, we shall assume 
the nanoparticle to have spin-degenerate single-electron energy levels
$|\psi_a\rangle$ with energy $E_a$, and a fully (doubly)
occupied Fermi level ($|\psi_F\rangle$) at energy $E_F$ [see Fig.
\ref{Fig_1}(a)].\cite{note-01}
Furthermore, by adopting $C_D/C_S>1$ the onset of tunneling corresponds to the 
injection of an electron from $S$ into $|\psi_{F+1}\rangle$, with a threshold bias voltage
given by $V^{th}_{S\rightarrow d}=(1/e)(C_{\Sigma}/C_{D})(U+E_{F+1}-E_F)$. As 
the number of electrons inside
the grain increases by one, the single-electron energy levels of the grain
are renormalized upwards by $U$, within the constant interaction
approximation\cite{ullmo_PRB_2001} [see Fig. \ref{Fig_1}(a)]. Therefore, the
subsequent tunneling processes that can take place are: (i)
return of the additional electron from level
$|\psi_{F+1}\rangle$ to $S$ or its transmission to $D$; or (ii) tunneling to $D$ of one of the
electrons populating the single-particle levels whose energy satisfies
$E_a+U\ge\mu^{D}(V^{th}_{S\rightarrow d})$. In the latter case, once the
electron is ejected from the grain, the remaining electrons are left in
an excited state ($|X_1\rangle$) that corresponds to creating one electron-hole ($eh$) pair
on the ground-state electronic configuration ($|G\rangle$) of the nanoparticle; as
exemplified in Fig. \ref{Fig_1}(a). Note that there are {\em different} possible 1eh-pair
electronic excitations depending on which electron is ejected out.
(We suppose that no shake-up\cite{mahan_book} occurs due to the creation of the 1eh-pair excitation.) The
slow electronic relaxation within the  grain, compared with the bottleneck tunneling rate (recall the
assumption that $\Gamma/\Gamma_r \gg 1$) makes it possible to generate another family of excitations, namely, 2eh-pair
excitations [see Fig. \ref{Fig_1}(b)]. The latter, however, require {\em two} consecutive
tunneling events for their creation. It should be noted that the number of possible 2eh-pair
states is greater than that of 1eh-pair states.

It is important to notice that the multi-eh-pair excitations only arise as a 
consequence of {\em multiple} sequential tunneling, as described above.
Hence, excited states are likely to exist only if $\Gamma_r$ is many times {\em smaller} than $\Gamma$.
Within this context, it is convenient to introduce the concept 
of {\em generations} of excitations: We define a {\em generation} to be the set 
of electronic configurations that contain a specific number of eh-pair excitations due to
consecutive electron tunneling  in and out of the grain. Hence, generation $n$ ($G\#n$) contains
the subset of $n$ eh-pair excitations that arise after $n$ such consecutive pairs of tunneling
events.  Figure \ref{Fig_1}(b) shows examples of members of generation 1 and generation 2, 
indicated by $G\#1$ and $G\#2$.
It should be clear that the generations constitute a {\em hierachy}: It is not possible to have an 
element of generation-$n$ without having generated previously elements of its $n-1$, 
$n-2$, $n-3$, $\cdots,1$ ancestors. Hence, in order for $n$ eh-pair excitations to occur with
significant probability the bottleneck tunneling time (${\Gamma}^{-1}$)  should be nearly $2n$ times {\em smaller} than
the electronic relaxation time (${\Gamma}_{r}^{-1}$). 
Finally, the applied bias voltage determines the maximum
number of generations that may be present in a device, as it
constrains the number of available energy levels:  The maximum
possible number of eh pairs is $2n$ if $n$ levels are
available in the grain for tunneling; therefore,  at most $2n$
generations might exist.

To find the {\em total} number of excitations present in generation $n$ 
we note that, {\em contrary} to the case in closed electronic
systems, the tunneling-induced  electron-hole pair excitations
{\em may have} a net total spin  polarization as large 
as $S_z=S^e_z+S^h_z$; where $S^{e(h)}_z$ is the largest
possible total spin polarization of the excited electrons
(holes).  This leads to a high multiplicity of the electronic 
configurations that may be present in generation $n$, as these
configurations result from exhausting all  the possible
electron and hole arrangements once the restriction $S_z=0$ is
waived. If $N$ ($M$) orbitals are available to accommodate the
excited electrons (holes), the total number of electronic
configurations up to generation $n$ is:
$\sum^{n}_{i=0}\,\biglb(\,^{2N}_i\bigrb)\biglb(\,^{2M}_i\bigrb)$; 
where $2n \leq N$ and the factor of 2 accounts for spin
polarization.

The above mechanism for generating electronic excitations within a metallic grain in the
Coulomb blockade  regime, and the subsequent identification of a hierarchy for these
excitations constitutes a generalization of the AWARTi model. In order to invoke this mechanism as
a possible interpretation of experimental tunneling spectroscopy data, one has to realize
that the one-electron levels in the ground-state of the grain are not spectrally
rigid.\cite{agam_PRL_1997,mahan_book}  In particular, a net electronic interaction due to the
redistribution of the charge within the {\em neutral} grain due to the excitation
renormalizes the one-electron levels: The ground-state single-electron  energy level structure
$(|\psi_a\rangle,E_{a})$ changes to
$(|\psi_a\rangle^{n}_{X_{n}}\,,E^{n}_{aX_{n}})$, with $|X_{n}\rangle$ an excited state in 
generation $n$. Figure \ref{Fig_1}(c) schematically shows this renormalization in a 
particular case (see Fig. \ref{Fig_2} also).   In general, for an excitation with a given
number of {\em eh} pairs, the renormalization  of level $|\psi_a\rangle$ depends on the
the specifics of the excited state, and therefore exhibits {\em fluctuations}; see
Fig. \ref{Fig_2}.  This is due to the stochasticity of the wavefunctions that is imposed by the 
disorder present in the nanograin.\cite{brody_RMP_1981} Random
matrix theory offers a simple generic approach to modeling this fluctuating renormalization 
of the one-electron levels.
In the presence of single-electron energy level renormalization the tunneling current $I(V)$, at bias voltage $V$,  is 
determined by a complex master equation that contains the kinetics of all 
the allowed electron tunneling transitions---including also the renormalized energy levels. Those are given 
by the following Coulomb blockade threshold equations: $\mu_S(V) \ge E^{n}_{aX_{n}} + U$ (for events involving 
{\em unoccupied} levels in the grain) and 
$\mu_D(V) \le E^{n}_{aX_{n}} + U$ (for events involving {\em occupied} levels in the grain); here $n=0$
represents the ground-state energy levels ( $E^{0}_{aX_{0}}=E_a$). Hence, each time a new tunneling channel is opened
a peak in the differential conductance is expected (see below, Sec. \ref{numerics_Al}). 

In summary, we have so far introduced the mechanism that creates electronic excitations,
recast these excitations as a  hierarchy of generations, stated that 
these excitations renormalize the single-electron energy levels in a complex manner, and
discussed the effect of the  renormalization on the tunneling spectroscopy of the grains. In
the following section we present a mean-field theory that can be used to calculate the renormalized
one-electron levels of the excited nanoparticle.

%%  how we incorporate the excitations %%
%%  THE MODEL %%
%
\section{Mean-field theory of the excited quasi-particle states}
\label{mean-field_model}

%%%
%%
%
% Ground-state electronic energy levels
%%
%%%

A neutral metallic grain with  
$n_0$ valence electrons in its ground-state ($|G\rangle$) is the starting point. 
We then adopt a quasi-particle picture consisting of electronic levels
$|\psi_a\rangle$ with energy $E_a$, and take as $|G\rangle$ the
lowest energy $n_0$-electrons configuration [see Fig. \ref{Fig_1}(a)]. 
%
%%%%%%%%        
% Hamiltonian
%%%%%%%%%
%
By implementing a tight-binding Hamiltonian, $E_a$ and $|\psi_a\rangle$ are calculated
taking into account important microscopic features of the nanograin: geometry, structural disorder, and surface
chemistry.\cite{narvaez_PRB_2002} 
In a multi-orbital ({\em s}, {\em p}, and {\em d}) representation, and 
neglecting spin-orbit coupling, the Hamiltonian is:

\begin{equation}
{\cal
H}=\sum_{i,\alpha}\varepsilon_{i}^{\alpha}c^{\dagger}_{i\alpha}c_{i\alpha}+
\sum_{i,j,\alpha,\alpha'}
{\cal T}_{{j\alpha},{i\alpha'}}\left(c^{\dagger}_{i\alpha'}c_{j\alpha}
+c^{\dagger}_{j\alpha}c_{i\alpha'}\right).
\label{tb_hamiltonian}
\end{equation}

\noindent $i$ ($j$) labels the lattice
site $\vec{R}_i$ $(\vec{R}_j)$ of atom $i$ $(j)$,
$c^{\dagger}_{i\alpha}$ $(c_{i\alpha})$ creates (destroys) an
electron on site $i$, while
$\alpha$ $(\alpha')$ indicates the {\em s}, {\em p} or {\em d} orbital. 
$\varepsilon_{i}^{\alpha}$ and ${\cal T}_{{j\alpha},{i\alpha'}}$
are the Slater-Koster (SK) on-site and hopping
parameters.\cite{slater_PR_1954,note-42}  

%
% morphology 
%
The metallic nanoparticles that are probed by electron tunneling spectroscopy 
are passivated with a thin Al-oxide layer and 
{\em buried} between two massive electrodes.
Hence, to extract data about 
the morphology of these nanograins is quite difficult.
However, their size---several hundreds of atoms---makes it reasonable to model their 
structure as  a crystaline metal {\em core}, and a disordered superficial ({\em shell})
region that corresponds to the  metal-oxide interface. The disorder present in the surface arises
from the chemistry of the oxide layer in combination  with surface reconstructions that may occur
during growth.\cite{note-40} 
%
% model nanoparticles
%
The model nanoparticles considered here result from truncating a fcc lattice to a volume ${\cal V}$ 
in a disc or hemisphere geometry.\cite{note-41}
In this case, the coordination numbers of atoms at the
surface of the nanoparticle differ from the coordination number of the bulk
fcc lattice. This is used to establish a criterion that defines
the surface of the particle, and distinguishes between core and shell.\cite{narvaez_PRB_2002}
Once the surface atoms are determined we {\em randomly} 
choose $50\%$ of those sites to represent O 
while the other sites correspond to Al.  This randomness is the only source of surface disorder we
consider.

%
% Oxide coating
%
We adopt as SK parameters in Eq. (\ref{tb_hamiltonian}) for the atomic sites in the core of the nanograin 
those parameterized by Papaconstantopoulos.\cite{papaconstantopoulos_book_1986}
The aluminum and oxygen sites in the metal-oxide
interface (shell), on the other hand, have different SK parameters 
as a consequence of charge transfer from {\em Al} to {\em O}.
To find the on-site energies for the {\em charged}
oxygen and aluminum atoms in the oxidized shell
we combine (i) the Mulliken-Wolfsberg-Helmholz (MWH) molecular-orbital
approach\cite{mcglynn_book} with (ii) results of classical molecular dynamics simulations
by Campbell and co-workers\cite{campbell_PRL_1999} that show the metal-oxide
interface at the nanometer scale is mainly constituted of intercalated $O^{-1/2}$ and $Al^{+1/2}$. 
The degree of charge transfer $\nu$ determines the on-site energies via the MWH theory:
The molecular orbital ($\mu$) energies of atom $m$ are empirically parameterized as 
$E^m_{\mu}=E^m_{\mu}(\nu)=-(A^m_{\mu}{\nu}^2+B^m_{\mu}\nu+C^m_{\mu})$;\cite{mcglynn_book} where $\nu$ 
is the excess valence charge: $\nu=(-)1/2$ for Al (O). 
Hence, the SK on-site energies for the atomic sites {\em in the oxide layer} are given by
${\varepsilon}^{\alpha}_{Al(O)}=-(A^{Al(O)}_{\alpha}/4+(-)B^{Al(O)}_{\alpha}/2+C^{Al(O)}_{\alpha}-\Delta)$,
with $\Delta=1.289Ry$ an off-set energy such that $E^{Al}_s(0)={\varepsilon}^s_{Al}$, and
the parameters $A^m_{\alpha}$, $B^m_{\alpha}$, and $C^m_{\alpha}$ for $m=Al,O$ those 
of Ref. \onlinecite{mcglynn_book}.
Finally, we determine the nearest neighbor {\em O-Al} hopping SK parameters by
assuming  an average separation between oxygen and
aluminum atoms in the oxide of $d_{Al-O}=1.8${\AA},\cite{campbell_PRL_1999}
applying Harrison's model to obtain the two-center transfer integrals
\cite{Harrison_book_1980} and then transforming them to find the SK
hopping parameters.\cite{papaconstantopoulos_book_1986}
By taking $d_{O-O}=3.0${\AA}$\,$\cite{ansell_PRL_1997} as average separation between {\em O} atoms the 
above procedure leads to the nearest neighbor {\em O-O} hopping parameter. 
We keep the SK hopping parameters for $Al^{+1/2}$-$Al^{+1/2}$ and 
$Al^{+1/2}$-$Al$  the same as those for {\em Al-Al}.

%%%
% Excited single-electron energy states
%%%

Let us now assume that
electronic excitations of the $n_0$ electrons within the grain are present
due to the mechanism discussed in Sec \ref{mechanism}. These excitations can be identified
according to the number of electron-hole pairs generated over the
ground-state
$|G\rangle$. A generic $n$-electron-hole ({\em n-eh}) pair excitation is given by:
$|X_{n}\rangle=|k^{<}_{1}k^{<}_{2}\cdots k^{<}_{n};k^{>}_{1}k^{>}_{2}\cdots k^{>}_{n}\rangle=(\Pi^{n}_{l=1}c^{\dagger}_{k^{>}_{l}})
(\Pi^{n}_{m=1}c_{k^{<}_m})|G\rangle$, where
$k^{>}_{l} $($k^{<}_{m}$) labels an empty (occupied) quasi-particle state
in the ground-state, and $c^{\dagger}_k$ ($c_k$) creates (destroys) an electron in state
$|\psi_k\rangle$. As mentioned above, the excitation changes the electronic charge density
within the grain from
$\rho_G(\vec{R})$, in the ground-state, to
$\rho_{X_{n}}(\vec{R})=\rho_G(\vec{R})-e[\sum^{n}_{l=1}
|\psi_{k^{>}_{l}}(\vec{R})|^2-
\sum^{n}_{m=1}|\psi_{k^{<}_{m}}(\vec{R})|^2]$ where $\vec{R}$ indicates an atomic site inside
the grain, and $|\psi_{k}(\vec{R})|^2$ is the wavefunction amplitude of
state $|\psi_k\rangle$ at site $\vec{R}$. Therefore there is an induced ({\em bare})
charge density in the grain given by:
$\delta\rho_{X_{n}}(\vec{R})=\rho_{X_{n}}(\vec{R})-\rho_G(\vec{R})$ which is
{\em screened} by the electrons present in the metallic grain.  
This {\em screening} can be modeled at different levels of complexity.\cite{blanter_PRL_1997} 
Here we implement for simplicity a {\em static}
screening represented by an effective Thomas-Fermi (TF) dielectric constant
$\epsilon^{TF}_q=1+(q_{TF}/q)^2$ ($q^2_{TF}=4(3\pi^{5}a_B^3{\cal N})^{-1/3}$ and $q=2/a_{Al}$ 
with $a_{Al}=0.405${\AA} and ${\cal N}=0.181${\AA}$^{-3}$ the bulk Al lattice parameter 
and electronic density, respectively; $a_B=0.529${\AA} is the atomic Bohr radius).\cite{ashcroft_book}
Within this approach the net (screened) induced charge is:
$\delta\rho_{X_{n}}(\vec{R})/\epsilon^{TF}_q$. 
The choice of the the wave vector magnitude
$q$ that enters the TF screening is due to the typical length
scale on which the electron wavefunction changes within the
grain.\cite{narvaez_PRB_2002b} 

To investigate the effect of the electronic excitations $\{|X_{n}\rangle\}$ in the energy spectra of
the grain, we extend our tight-binding model by setting the on-site energy of
orbital
$\alpha$ in atomic site
$\vec{R}_j$ to:
$\varepsilon^{X_n}_{j\alpha}=\varepsilon_{j\alpha}+\Sigma_{j\alpha}$. 
Here, 

\begin{equation}
\Sigma_{j\alpha}= U_{j\alpha}+
 \frac{e^2}{\epsilon^{TF}_q}
\sum_{i\neq j}\frac{\delta\rho_{X_{n}}(\vec{R}_i)}{|\vec{R}_i-\vec{R}_j|}
\label{on-site}
\end{equation}

\noindent is the renormalization energy introduced by the
Coulomb interaction due to the electronic excitation.
Within the empirical WHM approach discussed above,\cite{mcglynn_book} 
$U_{j\alpha}=(2A^{j}_{\alpha}q_j+B^{j}_{\alpha})e\delta\rho_{X_{n}}(\vec{R}_j)/{\epsilon^{TF}_q}$, and
accounts for  the {\em excess} charge $[-e\delta\rho_{X_{n}}(\vec{R}_j)/{\epsilon^{TF}_q}]$ present in
atomic site
$\vec{R}_j$ due to the redistribution of charge inside the nanograin; $q_j$ is the ground-state charge
present in site {\em j}.    The second term on the right hand
side of Eq. (\ref{on-site}) is the off-site Hartree contribution.\cite{note-12}
It should be noted that in the present model we do not consider 
any exchange effect---fine structure---that may distinguish between different electron-hole spin configurations.
Furthermore, we do not include scattering among different excitations $\{|X_{n}\rangle\}$.
This approximation should be valid for the low-lying electronic excitations.\cite{altshuler_PRL_1997}

%%	 what we investigated and what we found %%
%
\section{Case study: Aluminum nanograins}
\label{numerics_Al}

We now address the effects of electronic excitations on the
energy spectra of metallic nanograins by presenting detailed results
for the response 
of the single-electron energy levels to the creation of the lowest-lying excitations 
that are accessible at low applied bias voltage in a two-terminal device 
containing a disc-shaped Al nanograin coated with Al oxide. The grain's volume
${\cal V}=13.4\,nm^3$, the drain-source capacitance ratio $C_D/C_S=1.6$, and the charging
energy $U=4\delta$;\cite{note30} here $\delta=(4 E^{Al}_F/3 {\cal N}){\cal V}^{-1}=6.3\,meV$  is the 
average energy level spacing---predicted by a
particle-in-a-box model of the grain---around the Fermi energy of bulk Al $\left(E^{Al}_F\right)$. We consider 
three representative grains---{\em A}, {\em B}, and {\em C}--- that differ in the specific realization
of the disorder in their oxide coats.

\subsection{Ground state and excited energy levels}

The ground-state ($|G\rangle$) single-electron energy structure of nanograin {\em A} is shown in  Fig. \ref{Fig_2}(a); around its Fermi 
energy ($E_F$), and for a particular realization of disorder.\cite{note_0} 
Figure \ref{Fig_2}(a) also shows the renormalized energy spectra for generation 1 that arise by reaching the threshold voltage 
$V^{th}_{S\rightarrow d}=V^{th}_1$ to inject an electron into level $|\psi_{E_F+1}\rangle=|1\rangle$; the spectra 
are labeled by the particular 1eh electronic excitation that is involved. 
For this device at the threshold bias voltage, the
resulting singly charged nanoparticle can decay by emitting an electron from any of seven different 
single-electron spin-degenerate {\em orbitals}
into the drain contact. As previously discussed, the different
excitations renormalize the one-electron energy levels differently;  and fluctuations in the magnitude of the
energy renormalization, for a given level, are visible.  It should be noted that the renormalization of the 
energy levels is {\em smaller} than the {\em average} energy spacing of the ground-state single-electron 
energy levels:\cite{note31}
$\langle\delta\rangle \simeq 6.9\,meV$. Furthermore, for each {\em excited} state the average spacing between
one-electron energy levels differs  from $\langle\delta\rangle$ by less than $2\%$.
Increasing the bias voltage to $V^{th}_2$, at which it becomes possible for an electron to tunnel into level
$|\psi_{E_F+2}\rangle=|2\rangle$,
leads to {\em new} 1eh-pair excitations.  
Figure \ref{Fig_2}(b) shows in detail---within an energy
interval of $3\delta/5$---the excited single-electron levels around $E_{F+1}$ and $E_{F+2}$ that 
result from generation 1. 
For comparison, the box in the lower panel of Fig. \ref{Fig_2}(b) shows
the renormalized energy levels for generation 2; only for $E_{F+1}$, and at $V^{th}_1$.
Clearly, the fluctuations in the renormalization of the energy levels
depend {\em strongly}  on the details of the excitation(s) that are involved.

As mentioned above, disorder imparts a stochastic nature to
the confined electronic wavefunctions, thus dramatically affecting the
renormalization  fluctuations. The latter are visible in Figure \ref{Fig_3}, where the ground-state and 
excited electronic energy levels for different disorder realizations are shown: While for grain {\em A} the 
spectra show one of the renormalized levels of $E_{F+1}$ and $E_{F+2}$ visibly 
separated from the others, for all these 1eh-pair excitations, 
this is 
not the case for {\em B} and {\em C}; see Fig. \ref{Fig_3}. 
In particular, for grain {\em C}, the renormalized energy levels corresponding to $E_{F+2}$, at $V^{th}_2$, superpose with the 
renormalized levels for $E_{F+3}$.
It should be noted that the number of excitations is similar for the different nanograins.
This results from the combined effect of (i) adopting the same drain-source capacitance ratio ($C_D/C_S=1.6$) and 
charging energy ($U=4\delta$) for all the grains, and (ii) 
having the volume (${\cal V}$) fixed while changing the disorder from grain to grain. The latter gives nearly  the same {\em average} 
number of single-electron energy levels in an energy interval of {\em several} $\delta$ for each grain.\cite{narvaez_PRB_2002} 
In particular, the number of available orbitals for generating excitations, at $V^{th}_1$, 
is roughly: $int\{[U+E_{F+1}-E_{F}][1+(C_D/C_S)^{-1}]/\delta\}=int[6.5+1.625\,(E_{F+1}-E_F)/\delta]$, 
where $int(x)$ is the integer part of $x$.

In conclusion, the response of the electronic structure to the excitations exhibits strong
dependence on the energy levels involved in the excitation, 
and disorder realization.

\subsection{Differential conductance}
Having calculated the renormalized energy levels let us turn now to the implications for the differential
conductance. We concentrate here on finding the energies---or, equivalently, bias voltage---at which
tunneling resonances  should be present rather than performing an actual calculation of the $dI/dV$
spectrum. The latter is beyond the scope of this work as it would require to solve a complicate master
equation describing  the kinetics of the tunneling transitions far from equilibrium.

In general, resonances in $dI/dV$ arise when new channels for tunneling open as the bias voltage is
swept. In the Coulomb blockade regime, three kinds of tunneling resonances are expected in the 
spectra of metal nanograins: Direct, charge-fluctuation,\cite{narvaez_PRB_2002b} and
non-equilibrium.\cite{agam_PRL_1997} The first two are present regardless of the ratio between the
electronic relaxation and bottleneck tunneling rate, therefore they are regarded as {\em equilibrium} 
resonances. Direct resonances generally correspond to tunneling transitions directly into (out of) an
electronic  energy level in the ground state of the nanograin.   The charge-fluctuation resonances that
are present at low bias voltage arise from having a finite  probability for the nanograin to be (on
average) negatively charged---excess of one electron.\cite{narvaez_PRB_2002b}  These resonances are also
non-trivially affected by non-equlibrium effects (see below), however, they were {\em not included} in 
the original AWARTi model. Here, we generically label them $Q^+$. 

Figure \ref{Fig_4} summarizes the bias voltages ($V$), presented in units of energy after converting to $eV$, 
at which tunneling resonances should be present in the $dI/dV$ spectra of grains
{\em A}, {\em B}, and {\em C}; at low bias voltage. Resonances appear in groups: three in {\em A}, 
and two in {\em B} and {\em C}; only $G\#0$ (ground-state) and $G\#1$ are shown.
In these grains the Fermi level  is doubly occupied, as mentioned in Sec. \ref{mechanism}; furthermore, 
the adopted capacitance ratio $C_D/C_S$ is such that
in these devices the onset of tunneling takes place when an electron is injected from the source electrode into the grain.
Hence, the first resonance in $dI/dV$ appears when the bias voltage reaches $V^{th}_{1}$.
Upon increasing $V$, an additional {\em non-equilibrium} 
tunneling resonance appears in the spectrum each time a renormalized energy value of $E_{F+1}$ is reached. 
Further increase in the bias voltage results in new equilibrium and non-equilibrium resonances that arise from tunneling into 
$E_{F+2}$ (at $V^{th}_{2}$) and its renormalized energy levels, respectively.  
Non-equilibrium resonances appear as satellites 
of the main, equilibrium, resonances due to tunneling into $E_{F+1}$ and $E_{F+2}$.
This leads to group 1 and 3 in {\em A}; 1 and, partially, group 2 in {\em B}; and  group 1 and 2 in {\em C}.
It should be noted that the satellites involving level $E_{F+2}$, however, do not systematically appear
above the equilibrium resonance. The latter would make the experimental identification of the equilibrium resonance 
difficult. Moreover, the number of such satellite resonances is 
{\em bigger} than in the neighborhood of $V^{th}_1$ as the accessible 1eh-pair excitations at $V^{th}_2$ involve both levels 
$E_{F+1}$ and $E_{F+2}$.
Finally, at bias voltages greater that $V^{th}_1$ the finite probability to have the nanoparticle 
{\em negatively} charged leads to charge-fluctuation resonances\cite{narvaez_PRB_2002b}
(equilibrium and non-equilibrium satellites) that are responsible for group 2 ($Q^+$) in {\em A}, and part of group 2 in {\em B}.
However, no $Q^+$ resonances are present in grain {\em C} as a consequence of its electronic structure below
the Fermi energy.\cite{note-71}  
It is also noticeable that in grain {\em C} the clusters of resonances are less dense 
than in {\em A} and {\em B} although the number of 1eh-pair excitations
is nearly the same in all grains, as discussed above. This shows the high sensitivity of the tunneling spectrum to 
the response of the electronic energy level structure to eh excitations, and to the disorder present in the grain.

If we were to include in the above discussion of the $dI/dV$ spectrum the renormalized energy levels 
corresponding to $G\#2$ [see Fig. \ref{Fig_2}(b)] this would dramatically increase the number of non-equilibrium
resonances.  In fact, if the kinetics of the tunneling transitions is such as to admit a significant population of 
excited states then it  should be possible to achieve much higher spectral densities of tunneling resonances than observed
to date, simply by increasing the thickness of the tunnel barriers and hence the tunneling time relative to the relaxation
time of the eh pair excitations. 
This kinetics is determined by microscopic parameters of the grain, and the nature of the electron-phonon scattering within it; 
which strongly depend on the volume of the grain (see Ref. \onlinecite{arbouet_PRL_2003}) and disorder. 
Hence, there is no {\em a priori} (during the growth) control 
on the expected number of non-equilibrium resonances during charge transport in a device.
Nonetheless, tunneling experiments in gated devices (as those reported in Ref. \onlinecite{deshmukh_PRB_2002}), 
where the {\em charging energy} of the device may be changed with the applied gate voltage, 
should at least be able to address the increase of the number of
non-equilibrium resonances as a function of the available orbitals ($N$; see Sec. \ref{mechanism}) to 
generate the electronic excitations. 
Such observations may be contrasted
with the predicted number of excitations in generation $n$ as a function of $N$; see Sec.
\ref{mechanism}. However, the details of the  renormalization of the probed energy level
may hinder such a comparison. [see Fig. \ref{Fig_3} and \ref{Fig_4}, grain {\em C}, where the number of non-equilibrium 
resonances in the first cluster is smaller (nearly 1/2) than the number of accessible excitations, as a result
of having some renormalized levels (spectrally) below $E_{F+1}$. The latter implies that these excitations contribute to 
determine the tunneling current at $V^{th}_1$, rather than adding satellite resonances into the $dI/dV$ spectrum.]

To summarize, the high sensitivity of the ground-state and excited electronic energy spectrum to 
disorder is non-trivially inherited by the 
differential conductance spectrum: (i) The number of neutral tunneling resonances (equilibrium and non-equilibrium) strongly depend 
on the response of the electronic energy levels to the excitations. (ii) Charge-fluctuation resonances appear in clusters that, 
depending on disorder, can {\em overlap} with the neutral 
non-equilibrium resonances, as shown in Fig. \ref{Fig_4} (grain {\em B}). This would increase the 
complexity of the tunneling spectrum and make the interpretation of experiments difficult.

\section{Summary}

In summary, we have presented a systematic study of the tunneling-induced electronic 
excitations in non-magnetic metallic nanograins: (i) We have discussed the mechanism that creates electronic excitations 
and classified them according to the resulting number of electron-hole pairs, which enabled us to 
introduce the concept of generations of excitations: A subspace of excitations with a specific number of electron-hole pairs.
The generations form a hierarchy, and the number of elements in a generation is determined by the applied bias voltage
and characteristics of the device. (ii) We have proposed a general microscopic 
mean-field model that can be used to calculate the quasi-particle 
energy levels in the presence of a tunneling-induced excitation. Based on this model, we have presented detailed results 
for the response of the electronic structure of ultra-small aluminum grains coated with Al oxide to these excitations. However,
the model should be applicable to most of the non-magnetic metallic nanograins currently probed in tunneling spectroscopy
experiments. Our results show that disorder present at the surfaces of the nanograins imparts a fluctuating character to the
renormalization of the single-electron energy levels due to the excitations, and that this renormalization is smaller
than the typical energy spacing between single-particle levels. 
Furthermore, the tunneling spectra of the nanograins consist of equilibrium and non-equilibrium resonances, which appear in
clusters whose structure is dramatically affected 
by the high sensitivity of the ground-state and excited electronic energy
structure to disorder.

We have also shown that if the nonequilibrium resonances discussed here and in previous work\cite{agam_PRL_1997} are present
in the tunneling spectra of metal nanoparticles at all, then their density in the spectra should vary {\em greatly} from sample
to sample. This is because (a) the density of nonequilibrium resonances increases rapidly with the number of
{\em generations} of excited states that are populated in the nanograin and (b) this number of populated generations is in turn
very sample dependent.

Finally, we have suggested that tunneling spectroscopy experiments in gated ultrasmall nanograins may be useful to probe the 
variation in the number of non-equilibrium resonances as a function of the number of available orbitals for tunneling.

\section*{Acknowledgments}

We are grateful to John W. Wilkins for his helpful comments.
Funding from U. S. DOE grant DE-FG02-99ER45795 made it possible to complete this research at The Ohio State University.
The work at Simon Fraser University---where this research started---was funded by the Natural Science and 
Engineering Research Council (NSERC), and the Canadian Institute for Advanced Research (CIAR).

%%%%%%%%%%%%%%%%%%%%%%%%%%%%%%%%%%%%%%%
%
%

%
% Fig. 1
%
\begin{figure*}
\caption{(a) Mechanism for creating electronic excitations, exemplified with the creation of a 1eh-pair excitation:
from the source ($S$) electrode, an electron tunnels into energy level $|\psi_{F+1}\rangle=|1\rangle$ of an originally neutral grain 
containing $n_0$ conduction electrons in its ground state $|G\rangle$, with a charging energy {\em U} as 
represented in the plot; the grain gets negatively charged ($n_0\rightarrow n_0+1$), and the 
single-electron energy levels are renormalized by an amount {\em U}; subsequently, an electron tunnels out---from, in this example,
level 
$|\psi_{F-1}\rangle=|\bar{1}\rangle$---to the drain ($D$) contact  
which leaves the nanoparticle in an electronic excited state: $|1;\bar{1}\rangle=c^{\dagger}_{1}c_{\bar{1}}{|G\rangle}$. In the 
latter notation, $c^{\dagger}_1$ $(c_{\bar{1}})$ creates (destroys) an electron in state $|1\rangle$ $(|\bar{1}\rangle)$. 
(b) $G\#1$ and $G\#2$ are representative elements of generation 1, containing 1eh-pair excitations, 
and generation 2, corresponding to 2eh-pairs; details in text.
$|11;\bar{1}\bar{3}\rangle=c^{\dagger}_{1}c^{\dagger}_{1}c_{\bar{1}}c_{\bar{3}}{|G\rangle}$ is pictorially present in this example. 
(c) Schematic representation of the renormalization of 
the ground-state single-electron energy levels after generating excitation $|1;\bar{1}\rangle$---see text. The renormalized energy levels
are shown as dotted dashes.}
\label{Fig_1}
\end{figure*}

%
% Fig. 2
%
\begin{figure*}

\caption{(a) Energy levels for grain $A$ (volume ${\cal V}=13.4\,nm^3$) around the Fermi energy
($E_F$) of the ground-state grain in units of $\delta=6.3meV$---the average electron-in-a-box energy level spacing---for 
the ground-state ($|G\rangle$, bold dashes) and 
different one electron-hole pair excitations ($|X_1\rangle=c^{\dagger}_kc_{k'}|G\rangle=|k;\bar{k^{\prime}}\rangle$; $k$ and
$k'$ indicate the (excited) electron and (excited) hole one-electron levels, respectively. Thin dashes.) 
that may be created after the applied bias voltage reaches the threshold for tunneling an electron into 
level $|1\rangle$ in the neutral grain from $S$: $V^{th}_1$.
(b) Details of the renormalization of levels $|\psi_{F+1}\rangle$ and $|\psi_{F+2}\rangle$ arising from the presence of 1eh-pair 
excitations at $V^{th}_1$ and $V^{th}_2$---bias voltage to tunnel an electron into $|\psi_{F+2}\rangle$ from $S$. The box marks the
renormalized energy of single-electron level $|1\rangle$ that results from 2eh-pairs excitations. Full scale corresponds to 
energy interval of $3\delta/5$.
}
\label{Fig_2}
\end{figure*}

%
% Fig. 3
\begin{figure*}
\caption{Ground state $|G\rangle$ (bold dashes) and excited single-electron energy levels (thin dashes) 
for different realizations of disorder---grains {\em A}, {\em B}, and {\em C}.
Excited states belong to $G\#1$ (1eh-pair excitations), at bias 
voltages $V^{th}_1$ and $V^{th}_2$ (see text and caption of Fig. \ref{Fig_2}). The renormalization of the energy levels shows a strong
sample (disorder realization) variation. Notice the different energy scales in the plots. $E_F$ is the Fermi energy of 
each grain, and 
$\delta$ is the same as defined in Fig. \ref{Fig_2}.}
\label{Fig_3}
\end{figure*}

%
% Fig. 4
\begin{figure*}
\caption{Applied bias voltages (converted to energy: eV) derived from the calculated 
ground-state and excited energy spectra 
at which tunneling resonances would appear in the $dI/dV$ spectrum
of the studied devices (see text). 
Only generations 0 ($G\#0$) and 1 ($G\#1$) are considered when finding the
energies for direct and charge-fluctuation (equilibrium and non-equilibrium) resonances.
Thin and bold (blue in on-line version) bars distinguish between energies 
associated with direct and charge-fluctuation resonances, respectively.
Different heights mark equilibrium and non-equilibrium resonance energies: 3/2 and 3/4 for the former, while 1 and 1/2 label the
latter. In the top panel---grain A---the symbol $Q^+$
indicates a cluster of charge-fluctuation resonances (see text, and Ref. \onlinecite{narvaez_PRB_2002b}). 
The * indicates nearly-degenerate values of $eV$. 
The ordinate scale is the same in {all} panels, however, notice the different range of bias voltages in each plot. 
$U$ and $\delta$ have the same meaning as in the previous figures.}
\label{Fig_4}
\end{figure*}


\begin{thebibliography}{100}
\bibitem{note-41} In this model, the geometry can be arbitrary. However, discs and hemispheres are the most
widely assumed geometries in the literature. On the other hand, our choice of an fcc lattice host is consistent with the
bulk lattices of most of the metals studied experimentally at present.
\bibitem{note-12} While more sophisticated accounts for the {\em screening} 
($\Sigma_{j\alpha}$) may be implemented the results derived from this 
simple model should be qualitatively correct.
\bibitem{note31} This spacing is calculated by considering 200 energy levels in the ground-state, symmetrically 
located around $E_F$.
\end{thebibliography}
\end{document}